# DeepVARMA: A Hybrid Deep Learning and VARMA Model for Chemical Industry Index Forecasting

Xiang Li and Hu Yang
School of Information, Central University of Finance and Economics.
*Corresponding author(s). E-mail(s): hu.yang@cufe.edu.cn;
Contributing authors: lxiang0729@163.com;

**Abstract:** Since the chemical industry index is one of the important indicators to measure the development of the chemical industry, forecasting it is critical for understanding the economic situation and trends of the industry. Taking the multivariable nonstationary series-synthetic material index as the main research object, this paper proposes a new prediction model: DeepVARMA, and its variants DeepVARMA-re and DeepVARMA-en, which combine LSTM and VARMAX models. The new model firstly uses the deep learning model such as the LSTM remove the trends of the target time series and also learn the representation of endogenous variables, and then uses the VARMAX model to predict the detrended target time series with the embeddings of endogenous variables, and finally combines the trend learned by the LSTM and dependency learned by the VARMAX model to obtain the final predictive values. The experimental results show that (1) the new model achieves the best prediction accuracy by combining the LSTM encoding of the exogenous variables and the VARMAX model. (2) In multivariate non-stationary series prediction, DeepVARMA uses a phased processing strategy to show higher adaptability and accuracy compared to the traditional VARMA model as well as the machine learning models LSTM, RF and XGBoost. (3) Compared with smooth sequence prediction, the traditional VARMA and VARMAX models fluctuate more in predicting non-smooth sequences, while DeepVARMA shows more flexibility and robustness. This study provides more accurate tools and methods for future development and scientific decision-making in the chemical industry.

## 1 Introduction

The chemical industry, as a vital pillar of the national economy, has been instrumental in propelling rapid economic development in China. Particularly since the advent of economic reforms and opening-up policies, significant strides have been made in the petrochemical sector, positioning China as a leading global player in this domain [1]. With the unveiling of the "14th Five-Year Plan," the chemical industry faces new developmental opportunities and challenges[2]. These challenges include intensified market competition, heightened environmental protection demands, and accelerated technological innovation[3]. To effectively navigate this evolving landscape, the chemical industry must continually innovate and transform to meet the increasingly complex domestic and international market dynamics. Future development efforts in the industry will need to prioritize technological innovation, green development practices, sustainability, and customer-centric approaches[4]. The industry is witnessing a diverse array of emerging trends, such as collaborations with central enterprises, state-owned enterprises, and private enterprises[5], along with high-level dialogues, aimed at securing a more prominent global market presence. In light of these trends, the chemical industry is actively exploring strategies to enhance its overall competitiveness and promote the transformation and sustainable development of the chemical sector[6][7].

From the current state of industry development, synthetic materials play a crucial role in the chemical industry. Synthetic materials are widely used in industries such as automotive, aerospace, construction, electronics, and healthcare, providing critical support for technological advancement and development in these sectors[4]. The prices, supply and demand, and market trends of synthetic materials are not only vital for the development of the chemical industry but also directly impact the production and market performance of downstream industries. Therefore, accurate forecasting of the synthetic materials index is of significant importance for strategic decision-making in the chemical industry and related industries within the industrial chain.

Currently, there is a diverse array of research on forecasting chemical industry indices. Hu Aimei et al. constructed ARIMA and GARCH models to forecast international oil prices and found that when oil prices experience significant fluctuations, the GARCH model outperforms the ARIMA model[8]. Traditional forecasting methods such as Autoregressive Integrated Moving Average (ARIMA) and Generalized Autoregressive Conditional Heteroskedasticity (GARCH) models have certain advantages in forecasting time series data but often exhibit limitations when facing complex market dynamics and nonlinear features[9]. With the advancement of data science, machine learning, and deep learning, neural network-based forecasting models, such as Long Short-Term Memory (LSTM), have demonstrated outstanding performance in time series forecasting. They can effectively capture long-term dependencies and nonlinear features in time series data[10]. Tang Jing et al. used LSTM neural networks to study the fluctuation status of the Bohai Sea thermal coal price index and accurately forecast its future trends. They utilized a mixed kernel density estimation method to explore the uncertainty of future trends in the coal price index, achieving a probabilistic assessment of the future price fluctuation range[10]. Wu Qun proposed an optimized grey model algorithm combining adaptive particle swarm optimization with multidimensional grey model to forecast energy demand, achieving the highest accuracy in prediction even when the original data is incomplete[11]. Liu Lixia et al. constructed a least squares support vector machine for predicting oil futures prices, and their prediction performance was significantly better than that of the Radial Basis Function (RBF) neural network. These studies demonstrate the diverse methodologies being employed in forecasting chemical industry indices, leveraging various models and algorithms to enhance prediction accuracy and effectiveness[12]. Yang et al. introduced a novel hybrid model based on Lasso and CNN, which exhibits enhanced capability in capturing complex relationships within time series data of gasoline prices. Compared to using Lasso or CNN models independently, the hybrid model significantly improves the accuracy and stability of predictions[13].

In conclusion, while statistical models may encounter limitations when dealing with nonlinear and non-stationary data, machine learning models may have limited ability to model interactions and dynamic relationships among multiple variables. Thus, the integration of statistical and machine learning models has emerged as a hot topic in the field, aiming to leverage the strengths of both approaches to achieve more accurate and stable predictions[14][15]. This paper focuses on non-stationary multivariate time series, specifically the synthetic materials index, and explores how to combine traditional statistical models with machine learning models to improve the accuracy and robustness of forecasting non-stationary correlated sequences. The proposed DeepVARMA algorithm integrates the statistical VARMA model with the machine learning LSTM model. It not only utilizes LSTM to extract long-term trends and nonlinear features from the time series data to obtain stationary residual sequences but also employs the VARMA model to capture interactions and complex dynamic relationships among the multivariate residual sequences. This combination leverages the predictive capabilities of LSTM on the original sequences and further refines the predictions through VARMAX on the residuals, enhancing the robustness of the forecast results. The dual prediction mechanism of the DeepVARMA model provides a more comprehensive description and prediction of non-stationary multivariate sequences such as the synthetic materials index. It provides a scientific basis for enterprise decision-making and government policy formulation, promotes the sustainable development and international competitiveness of the industry, and lays a solid theoretical and practical foundation for the development of the chemical industry.

The main contributions of this paper can be summarized as follows:

- To achieve the goal of capturing interactions between variables while extracting trends from time series data, this study proposes a multivariate time series forecasting algorithm based on DeepVARMA and its variants, DeepVARMA-re and DeepVARMA-en. This algorithm combines the Long Short-Term Memory (LSTM) model with the statistical model Vector Autoregressive Moving Average (VARMA), integrating long-term trend extraction with multivariate dependency information to optimize the forecasting of multivariate time series.

- Using a dataset of synthetic material indices from the chemical industry, comparative experiments were conducted to evaluate the effectiveness of the proposed algorithm. Experimental results demonstrate that, in the prediction of stationary multivariate sequences, DeepVARMA-en exhibits the best predictive performance. Conversely, in the prediction of non-stationary multivariate sequences, DeepVARMA demonstrates better overall prediction accuracy. Moreover, the proposed algorithm exhibits superior prediction stability across multivariate time series with different levels of stationarity compared to existing statistical models.

The organizational structure of the paper is as follows: Chapter 2 introduces relevant work; Chapter 3 presents the DeepVARMA-based multivariate time series forecasting algorithm and its variants; Chapter 4 evaluates and analyzes the proposed algorithm's performance; Chapter 5 provides a summary of the entire paper.

## 2 Related Work

Multivariate time series forecasting methods leverage correlations among multiple time series to enhance overall prediction performance. The key to forecasting multivariate time series lies in accurately capturing complex temporal patterns and dependencies among variables[15]. Vector Autoregression (VAR) is a commonly used statistical time series analysis method that treats multivariate time series as multidimensional stochastic processes. It builds upon the foundation of the single-variable Autoregressive (AR) model, assuming linear dependencies among variables[16]. VAR models describe stochastic processes linearly and are suitable for capturing common linear patterns among multivariate sequences, particularly applicable in modeling stationary sequences. Vector Autoregressive Moving Average (VARMA) models, an extension of the ARIMA algorithm, have found widespread application in various fields such as economics, finance, engineering, and social sciences[17]. Lütkepohl studied the application of VARMA models in predicting multivariate financial data, highlighting their ability to capture dynamic correlations among financial indicators, thus providing valuable insights for financial market forecasting[18][19]. In the domains of energy and manufacturing, Tsay utilized VARMA models to analyze multivariate forecasting issues in energy markets, revealing their high accuracy in predicting energy market dynamics[20]. This provides crucial references for optimizing resource allocation and enhancing production efficiency, thereby contributing to energy market planning and management. Despite their applications across diverse domains, these statistical time series analysis models assume linear temporal patterns and may not be suitable for modeling long-term dependencies.

Due to their outstanding performance in complexity and sequence feature representation, deep learning models are increasingly being employed in time series forecasting algorithms to capture complex temporal patterns and dependencies among variables[21]. Long Short-Term Memory (LSTM) networks, a variant of Recurrent Neural Networks (RNNs), were proposed by Hochreiter and Schmidhuber in 1997 to address the issue of vanishing and exploding gradients commonly encountered in traditional RNN models during training[22]. Over the past few decades, LSTM models have received extensive research attention and applications due to their exceptional performance in handling complex time series data[23]. Research by Zhao Hongke et al. on dynamic forecasting in the internet finance market[24], based on deep neural network structures, indicates that LSTM models demonstrate good performance in various time series prediction tasks across different domains[25][26]. The superior performance of LSTM models in time series forecasting can be attributed to their structural design, which incorporates memory cells and gate mechanisms capable of effectively capturing both long-term and short-term dependencies in time series data[27][28]. This enables LSTM models to perform well in predicting time series with complex dynamic behaviors and nonlinear features[29][30]. Moreover, LSTM has various variants, such as Gated Recurrent Units (GRUs) with simplified gate control mechanisms[16][31][32], phi-LSTM focusing on sequence phase information[33], and skip-LSTM explicitly connecting across time steps[34]. However, despite the strong performance of LSTM models in handling complex time series, there may still be limitations in modeling interactions among multiple variables.

## 2.1 VARMA

### 2.1.1 Model Expression

The VARMA model, short for Vector Autoregressive Moving Average Model, stands as a classical model in time series analysis. Initially introduced in 1975 by George E.P. Box and Gwynne M. Jenkins in their work "Time Series Analysis: Forecasting and Control," this model has remained a cornerstone in the field. It enables the simultaneous consideration of interdependencies among multiple variables, capturing their interactive effects. The VARMA model finds applicability in the analysis and forecasting of multivariate time series data[35][36]. Comprising two fundamental components, the VARMA model consists of the Vector Autoregressive (VAR) part and the Moving Average (VMA) part. The VAR component delineates the linear relationship between a time series and its past values. Each variable in the model can be predicted using the past values of other variables as well as its own historical values. On the other hand, the VMA component characterizes the cumulative impact of error terms (random disturbances) over a defined period in the past. It aids in capturing the stochastic fluctuations and non-systematic noise within the time series data.

Hannan and Deistle[37] give a general representation of the VARMA model:

$$y_t = A_1 y_{t-1} + A_2 y_{t-2} + \cdots + A_p y_{t-p} + \epsilon_t + B_1 \epsilon_{t-1} + B_2 \epsilon_{t-2} + \cdots + B_q \epsilon_{t-q} \quad (1)$$

Where $y_t$ is the m-dimensional vector of endogenous variables, $p$ represents the number of lags in the VAR part, $q$ represents the number of lags in the VMA part, $T$ is the time step. $A_1, A_2, \ldots, A_p$ and $B_1, B_2, \ldots, B_q$ is the $m \times m$ dimensional matrix of coefficients to be estimated, $(\epsilon_t | t = 1,2, \ldots, T)$ is the m-dimensional white noise with positive definite covariance matrix. The matrix expression form of Eq. (1) is as follows:

$$\begin{bmatrix} y_{1t} \\ y_{2t} \\ \vdots \\ y_{mt} \end{bmatrix} = A_1 \begin{bmatrix} y_{1t-1} \\ y_{2t-1} \\ \vdots \\ y_{mt-1} \end{bmatrix} + \cdots + A_p \begin{bmatrix} y_{1t-p} \\ y_{2t-p} \\ \vdots \\ y_{mt-p} \end{bmatrix} + \begin{bmatrix} \epsilon_{1t} \\ \epsilon_{2t} \\ \vdots \\ \epsilon_{mt} \end{bmatrix} + B_1 \begin{bmatrix} \epsilon_{1t-1} \\ \epsilon_{2t-1} \\ \vdots \\ \epsilon_{mt-1} \end{bmatrix} + \cdots + B_q \begin{bmatrix} \epsilon_{1t-q} \\ \epsilon_{2t-q} \\ \vdots \\ \epsilon_{mt-q} \end{bmatrix} \quad (2)$$

In addition, the VARMA model can be extended to include exogenous variables $x_t$ of the VARMAX model, which is generally expressed as follows:

$$y_t = \sum_{i=1}^{p} \Phi_i y_{t-i} + \sum_{k=0}^{s} \Gamma_k x_{t-k} + \sum_{j=0}^{q} \Theta_j \epsilon_{t-j} \quad (3)$$

Where $y_t$ is an m-dimensional endogenous variable vector, and $x_t$ is the n-dimensional vector of exogenous variables. $\Phi_i$ and $\Theta_j$ are the $m \times m$ coefficient matrices to be estimated for the autoregressive and moving average parts. $\Gamma_k$ is the coefficient matrix to be estimated for exogenous variables. $p$ represents the number of lags in the VAR part, $q$ represents the number of lags in the VMA part, $T$ is the time step, $B$ is the lag operator, $I$ is the identity matrix, and $(\epsilon_t | t = 1,2, \ldots, T)$ is an m-dimensional white noise with a positive definite covariance matrix.

### 2.1.2 Parameter Estimation

Given a multivariate time series $y_t$, the VARMA model expression can be written as:

$$\begin{cases} \Phi(B) y_t = \Theta(B) \epsilon_t \\ \Phi(B) = I - \sum_{i=1}^{p} \Phi_i B^i \\ \Theta(B) = I + \sum_{j=1}^{q} \Theta_j B^i \end{cases} \quad (4)$$

Where $y_t$ is an m-dimensional endogenous variable vector. $\Phi(B)$ and $\Theta(B)$ are respectively autoregressive (AR) and moving average (MA) polynomials. $\Phi_i$ and $\Theta_j$ are the $m \times m$ coefficient matrices to be estimated for the autoregressive and moving average parts. $p$ represents the number of lags in the VAR part, $q$ represents the number of lags in the VMA part, T is the time step, $B$ is the lag operator, $I$ is the identity matrix, and $(\epsilon_t | t = 1,2, \ldots, T)$ is an m-dimensional white noise with a positive definite covariance matrix.

The m-dimensional time series $y_t$ is a function based on past observations and the current error term, and it is assumed that the error term $\epsilon_t$ is zero-mean and the covariance matrix is Σ of the multivariate normal distribution, therefore, the $y_t$ the likelihood function can be expressed as:

$$L(\Phi, \Theta, \Sigma | y_1, \ldots, y_n) = \prod_{t=\max(p,q)}^{n} \frac{1}{\sqrt{(2\pi)^k \det(\Sigma)}} \exp(-\frac{1}{2}\epsilon_t^T \Sigma^{-1} \epsilon_t) \qquad (5)$$

where $n = m \times T$. $\epsilon_t = y_t - \hat{y}_t$ is the model residual term.

Maximum likelihood estimation is to find a set of parameter values such that the likelihood function is $L(\Phi, \Theta, \Sigma | y_1, \ldots, y_n)$ is maximized, and the objective function is the log-likelihood function, which can be expressed as:

$$logL(\Phi, \Theta, \Sigma | y_1, \ldots, y_n) = \sum_{t=\max(p,q)}^{n} (-\frac{k}{2}\log(2\pi) - \frac{1}{2}\log(\det(\Sigma)) - \frac{1}{2}\epsilon_t^T \Sigma^{-1} \epsilon_t)(6)$$

The optimal parameter estimates for the VARMA model can be obtained by maximizing the log-likelihood function.

## 2.2 LSTM

Long Short-Term Memory (LSTM) networks, an advancement over Recurrent Neural Networks (RNNs)[20], excel in handling long-term dependencies. They mitigate the issues of vanishing and exploding gradients present in traditional RNNs through gate mechanisms[38]. These mechanisms effectively capture and retain long-term dependencies within sequential data[39]. The internal recurrent structure of an LSTM model comprises four network layers, including output gates, input gates, and forget gates, along with a memory cell. These components collectively govern the forgetting and updating of outputs and memories[40]. The rough structure of the LSTM model is illustrated in Fig 1.

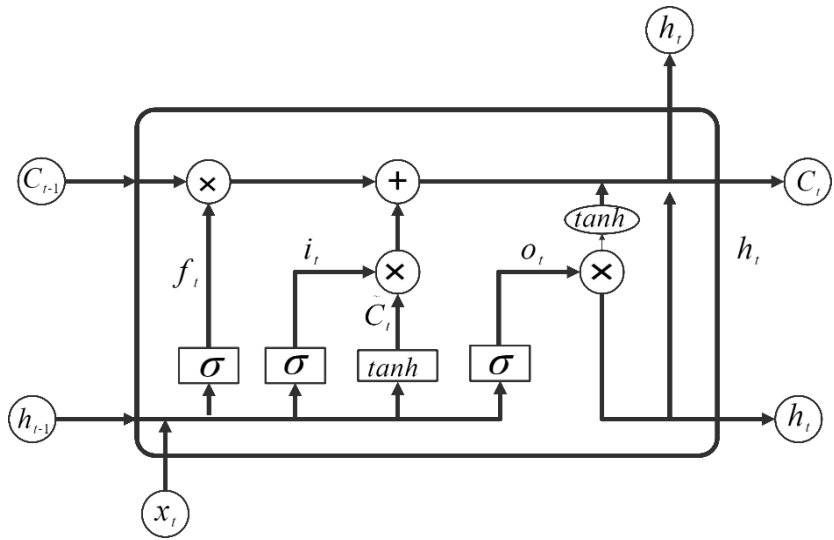

**Fig. 1** Diagram of LSTM loop structure

The LSTM model first determines the degree of forgetting $C_{t-1}$ through the forget gate. The sigmoid function serves as the activation function, transforming outputs into values ranging between 0 and 1. By employing the sigmoid function, $h_{t-1}$ and $x_t$ are converted into $f_t$, determining the extent to which information in the memory state is to be forgotten, where $f_t = 1$ indicates full retention and $f_t = 0$ indicates complete forgetting. Subsequently, the input gate decides which information to add to the memory state. The tanh function also acts as the activation function, converting outputs into values ranging between -1 and 1. Within the input gate, $h_{t-1}$ and $x_t$ are first transformed into $\tilde{C}_t$ using the tanh function, and then $h_{t-1}$ and $x_t$ are converted into $i_t$ using the sigmoid function. Finally, through the output gate, $C_t$ is transformed into $\tanh C_t$ using the tanh function, and $h_{t-1}$ and $x_t$ are transformed into $O_t$ using the sigmoid function.

## 3 Model Construction

In existing research, both LSTM and VARMA models exhibit their respective strengths and limitations. LSTM excels in handling complex time series data, particularly in capturing nonlinearity and long-term dependencies. However, LSTM's capability to model interactions and dynamic relationships among multiple variables is relatively limited. Conversely, VARMA demonstrates strong capabilities in modeling and forecasting multivariate time series data, yet it may encounter constraints when dealing with nonlinear and non-stationary data. By combining LSTM and VARMA models, it is possible to leverage the LSTM model's ability to handle complex time series data and the VARMA model's advantages in modeling multivariate interactions, thereby constructing a novel forecasting approach.

### 3.1 DeepVARMA-re

#### 3.1.1 LSTM Prediction Layer

Given a time series $y_t$ that represents observations at time *t*. This time series is modeled using a Long Short-Term Memory Network (LSTM) model to extract the trend portion of it. By training the LSTM model to learn the long term dependencies and trend changes in the time series[40], the current hidden layer state of the LSTM $h_t$ is determined by the current input $y_t$ and the hidden layer state at moment t-1 $h_{t-1}$ which are jointly determined by $h_t$ denotes the trend of the sequence prediction of the time series at time point *t* which is denoted as $\mu_t$ :

$$\mu_t = LSTM(y_t, h_{t-1}) \tag{7}$$

### 3.1.2 VARMA Prediction Layer

The LSTM layer has extracted the trend portion of the time series by subtracting the trend term from the original time series and calculating the sequence residuals, denoted as $e_t$:

$$e_t = y_t - \mu_t$$

$$= A_1(y_{t-1} - \mu_{t-1}) + \cdots + A_p(y_{t-p} - \mu_{t-p}) + \epsilon_t + B_1\epsilon_{t-1} + \cdots + B_q\epsilon_{t-q} \tag{8}$$

where $e_t$ denotes the non-trend components or irregular components that are not fully explained by the LSTM model. Next, a VARMA model is employed to model and forecast the residual component $e_t$.

Assuming that $p$ denotes the order of the autoregressive part and $q$ denotes the order of the moving average part, the $e_t$ of VARMA model modeling expression is as follows:

$$e_t = \sum_{i=1}^{p} \Phi_i e_{t-i} + \sum_{j=0}^{q} \Theta_j \epsilon_{t-j} \tag{9}$$

where $\Phi_i$ and $\Theta_j$ are respectively the coefficient matrices for the autoregressive and moving average parts, and $\epsilon_t$ represents the white noise error term of the model. After forecasting the residual $e_t$ using the VARMA model, the predicted value of the residual $e_t$ is obtained, denoted as $\hat{e}_t$.

### 3.1.3 Prediction Results Merge Layer

The final sequence prediction result, denoted as $\hat{y}_t$, is obtained by adding the trend component predicted by the LSTM model and the residual component predicted by the VARMA model.

$$\hat{y}_t = \mu_t + \hat{e}_t \tag{10}$$

This combined forecasting method integrates the strengths of both the LSTM and VARMA models. It effectively captures the trends in the time series while also providing accurate predictions for the residual component, thereby enhancing the accuracy of the forecasts.

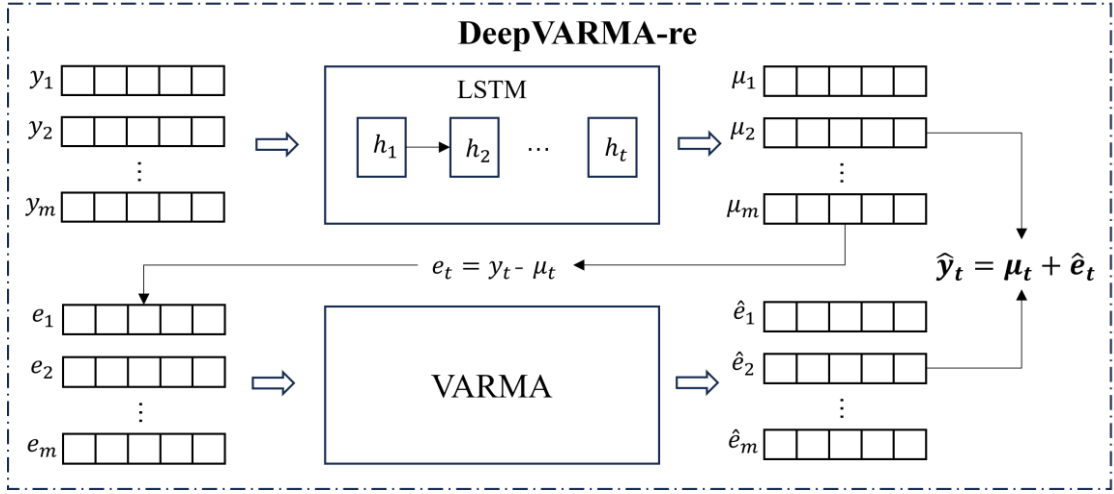

**Fig. 2** Structure of DeepVARMA-re

### 3.2 DeepVARMA-en

#### 3.2.1 LSTM Coding Layer

Given an n-dimensional exogenous variable $x_t$, the LSTM model learns the long-term dependencies and features of the exogenous variables, encoding them into a high-dimensional feature vector. The current hidden state $h_t$ of the LSTM is jointly determined by the current input $x_t$ and the hidden state $h_{t-1}$ at time $t$-1. $h_t$ represents the feature vector of the time series at time point $t$, denoted as $H_t$.

$$H_t = LSTM(x_t, h_{t-1}) \tag{11}$$

The LSTM model processes sequential data and extracts significant features, generating a fixed-length feature representation. These features serve as inputs to the VARMAX model.

#### 3.2.2 VARMAX Prediction Layer

The high-dimensional feature vectors encoded by the LSTM are used as exogenous variables in VARMAX for endogenous variable modeling and prediction, given the m-dimensional endogenous variable time series $y_t$ and the d-dimensional exogenous variable time series encoded by the LSTM model $H_t$, the expression of the VARMAX model is:

$$\begin{cases} \Phi(B)y_t = \Theta(B)\epsilon_t + \sum_{k=0}^{s} \Gamma_k H_{t-k} \\ \Phi(B) = I - \sum_{i=1}^{p} \Phi_i B^i \\ \Theta(B) = I + \sum_{j=1}^{q} \Theta_j B^i \end{cases} \tag{12}$$

where $y_t$ is an m-dimensional endogenous variable vector, and $H_t$ is the feature vector of exogenous variables encoded by the LSTM model. $\Phi(B)$ and $\Theta(B)$ are respectively autoregressive (AR) and moving average (MA) polynomials. $\Phi_i$ and $\Theta_j$ are the $m \times m$ coefficient matrices to be estimated for the autoregressive and moving average parts. $\Gamma_k$ is the coefficient matrix to be estimated for exogenous variables. $p$ represents the number of lags in the VAR part, $q$ represents the number of lags in the VMA part, $T$ is the time step, $B$ is the lag operator, $I$ is the identity matrix, and $(\epsilon_t | t = 1,2,\dots,T)$ is an m-dimensional white noise with a positive definite covariance matrix.

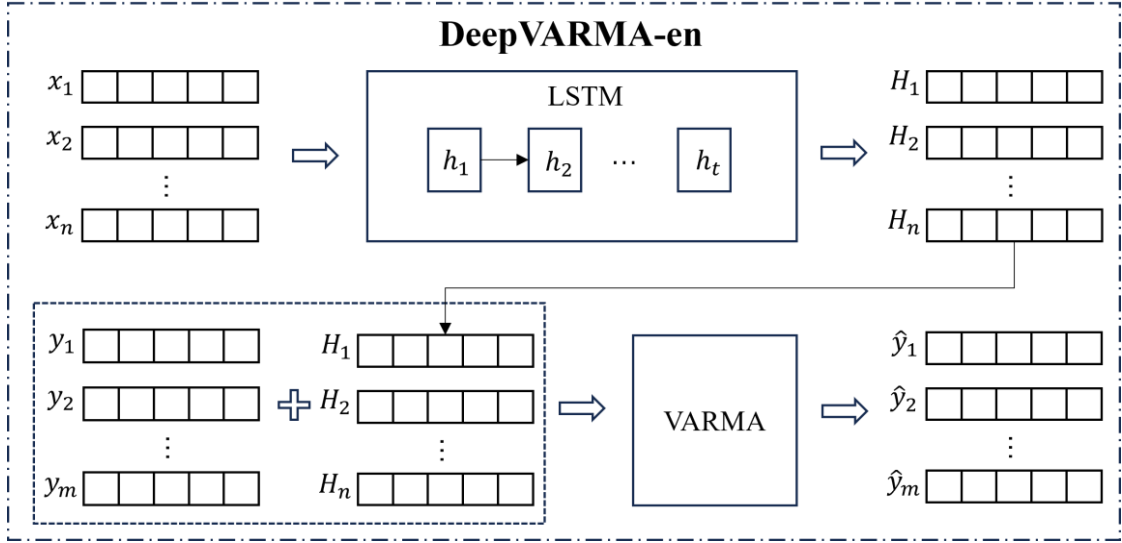

**Fig. 3** Structure of DeepVARMA-en

## 3.3 DeepVARMA

Based on DeepVARMA-re and DeepVARMA-en, we integrate the two models structurally to construct the DeepVARMA model. Firstly, we utilize the LSTM model to forecast endogenous variables and compute residuals. Secondly, we employ the LSTM model to encode exogenous variables. Subsequently, the encoded exogenous variables are input into the VARMAX model to forecast residuals of endogenous variables. Finally, we merge the predictions of endogenous variables from the LSTM model with the forecasts of residuals from the VARMAX model to obtain the ultimate prediction results. The following presents the principles and formula derivations of the DeepVARMA model.

### 3.3.1 LSTM Prediction Layer

Given an m-dimensional endogenous variable time series $y_t$, the prediction is made using LSTM model and the prediction result is obtained $\mu_t$, the LSTM model prediction expression is:

$$\mu_t = LSTM(x_t, h_{t-1}) \tag{13}$$

Then, the residuals between the true and predicted values are calculated and denoted as $e_t$:

$$e_t = y_t - \mu_t \tag{14}$$

### 3.3.2 LSTM Coding Layer

Given an n-dimensional time series of exogenous variables $x_t$, an LSTM model is employed to encode it. The LSTM model processes each step of the exogenous variables and ultimately obtains the encoded feature representation $H_t$. The encoding process of the LSTM model can be expressed as:

$$H_t = LSTM_{encoder}(x_t) \tag{15}$$

The encoded feature representation $H_t$ serves as the exogenous variable input for the VARMAX model.

### 3.3.3 VARMAX Prediction Layer

The encoded exogenous variable features are represented as $H_t$ from the LSTM serves as the exogenous variable input for the VARMAX model, which models and forecasts the residual sequence $e_t$ of the endogenous variables. The expression of the VARMAX model is as follows:

$$\begin{cases} \Phi(B)e_t = \Theta(B)\varepsilon_t + \sum_{k=0}^{s} \Gamma_k H_{t-k} \\ \Phi(B) = I - \sum_{i=1}^{p} \Phi_i B^i \\ \Theta(B) = I + \sum_{j=1}^{q} \Theta_j B^i \end{cases} \tag{16}$$

where $e_t$ is the m-dimensional endogenous variable residual vector, and $H_t$ is the vector of exogenous variable features encoded by the LSTM model. $\Phi(B)$ and $\Theta(B)$ are respectively autoregressive (AR) and moving average (MA) polynomials. $\Phi_i$ and $\Theta_j$ are the $m \times m$ coefficient matrices to be estimated for the autoregressive and moving average parts. $\Gamma_k$ is the coefficient matrix to be estimated for exogenous variables. $p$ represents the number of lags in the VAR part, $q$ represents the number of lags in the VMA part, $T$ is the time step, $B$ is the lag operator, $I$ is the identity matrix, and $\varepsilon_t$ is the model error term.

The VARMAX model models the dependency relationship between the residuals $e_t$ and the exogenous variable feature vector $H_t$, ultimately yielding the predicted residuals $\hat{e}_t$.

### 3.3.4 Prediction Results Merge Layer

The final prediction result $\hat{y}_t$ is obtained by adding the forecasted endogenous variable $\mu_t$ from the LSTM model and the predicted residuals $\hat{e}_t$ from the VARMAX model.

$$\hat{y}_t = \mu_t + \hat{e}_t \tag{17}$$

The DeepVARMA model integrates the nonlinear forecasting capability of LSTM models for time series with the multivariate linear forecasting capability of VARMAX models, thereby enhancing prediction accuracy and model performance.

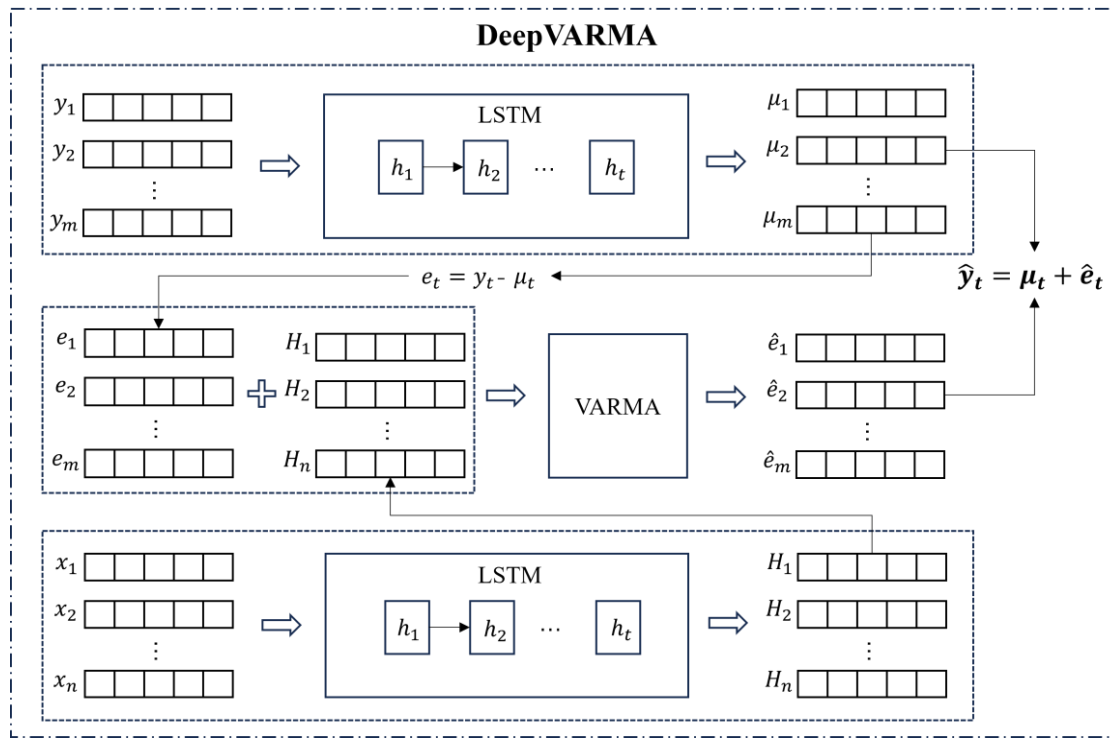

**Fig. 4** Structure of DeepVARMA

## 4 Experiments

### 4.1 Datasets

#### 4.1.1 Index Selection

The significance of synthetic materials in the chemical industry cannot be overstated. They constitute a vital component of modern industry, playing a crucial role in the development and innovation of various fields. In this study, data from the Synthetic Materials Index spanning from April 2, 2021, to April 1, 2024, is selected as the primary research focus. Specifically, this includes indices for synthetic fibers, plastics, and rubber. Additionally, external variables such as energy indices, exchange rates, and the price of wheel steel (transportation index) are chosen as exogenous variables influencing the dependent variables. The data is sourced from the Steel Union Data Center, the National Bureau of Statistics, and Wind Financial Terminal. The line graphs depicting each index series are illustrated in Figure 5.

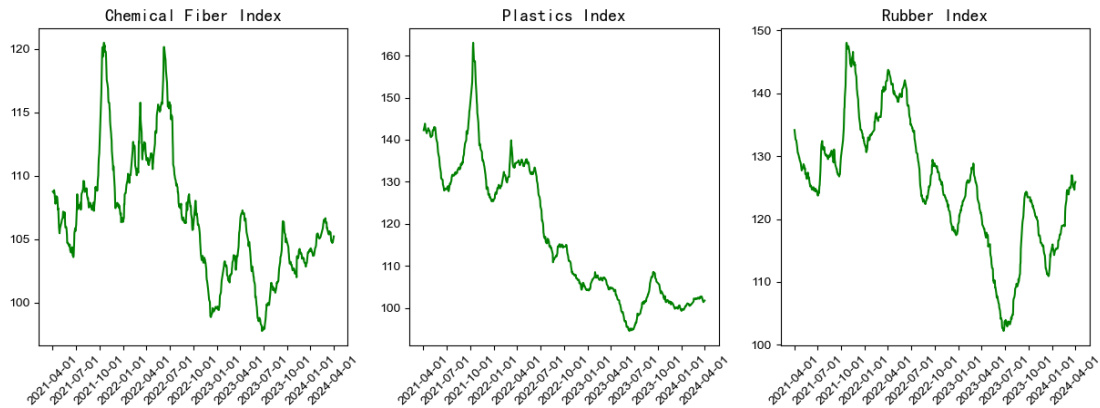

**Fig. 5** Trend of the composite index for each chemical material

The VARMA model is a multivariate time series model designed to capture the interdependencies and common patterns among multiple time series. Before using the VARMA model for modeling, examining the correlation between time series is a crucial step. By assessing the correlation between time series, one can better understand whether these sequences exhibit common trends, periodicity, or other correlated features. Drawing a heatmap of correlations to observe the strength of association between various indices, as shown in figure 6, reveals that the correlations between the indices are all close to 0.8, which is in line with the assumptions of the model.

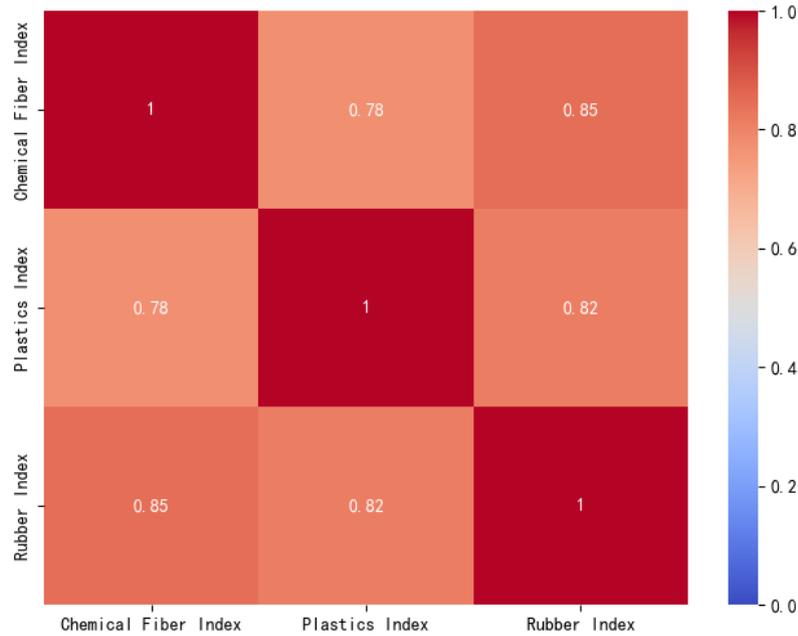

**Fig. 6** Correlation of the composite index of each chemical material

**4.1.2 Data Processing**

Before conducting experiments, missing data needs to be imputed, and logarithmic transformations need to be applied to exogenous variables. As this study aims to compare the predictive performance of different models under stationary and non-stationary conditions, it is necessary to test the stationarity of the data. Stationarity of a time series refers to the stability of its statistical characteristics (such as mean, variance, covariance, etc.) over time, implying that the properties of a stationary time series remain constant over time, and the shape of the series does not undergo significant changes over time. This study conducted stationarity tests on both the original series and the first-order differenced series, as shown in Table 1. The first-order differenced series are illustrated in Figure 7. It can be observed that the original series are non-stationary, while the first-order differenced series for each chemical index exhibit stationarity.

**Table 1** p-values before and after differencing for each chemical material composite index

|  | Original sequence p-value | First-order difference series p-value |
| --- | --- | --- |
| chemical fiber index | 0.214 | 0.000 |
| Plastics Index | 0.546 | 0.000 |
| Rubber Index | 0.335 | 0.000 |

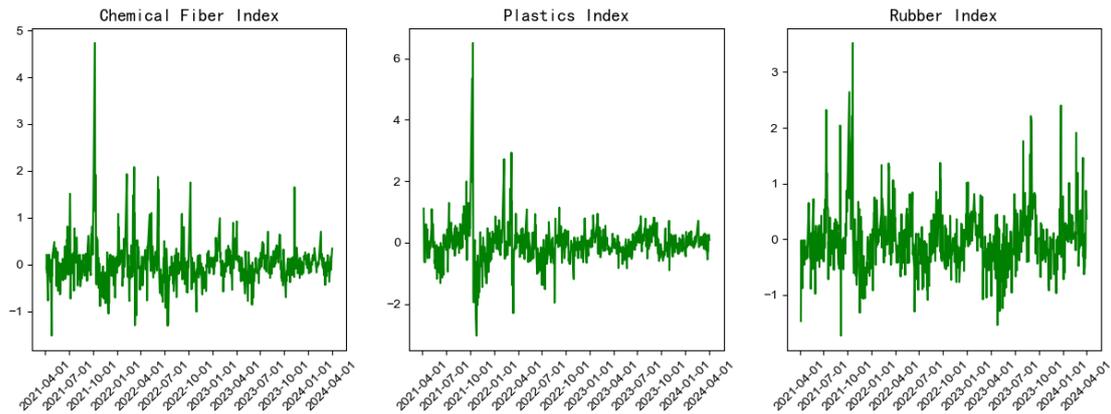

**Fig. 7** First-order difference series of each chemical materials composite index

### 4.1.3 Datasets Partition

The cleaned data are divided into training set, validation set and test set according to the ratio of 6:2:2, and finally, we get 449 pieces of data in the training set, 149 pieces of data in the validation set and 149 pieces of data in the test set. The descriptive statistics of the cleaned data are shown in Table 2.

**Table 2** Descriptive statistics

|  | Chemical Fiber index | Plastics Index | Rubber Index | Energy Index | Exchange Rates | Traffic Index |
|---|---|---|---|---|---|---|
| sample size | 747 | 747 | 747 | 747 | 747 | 747 |
| maximum values | 120.48 | 163.06 | 148.02 | 7.25 | 5.15 | 8.88 |
| minimum value | 97.76 | 94.54 | 102.21 | 6.31 | 4.61 | 8.66 |
| average value | 106.62 | 117.34 | 125.55 | 6.79 | 4.91 | 8.77 |
| standard deviation | 4.91 | 16.11 | 10.42 | 0.31 | 0.14 | 0.06 |
| skewness | 0.66 | 0.42 | -0.13 | -0.13 | 0.22 | -0.18 |
| kurtosis | 0.18 | -1.05 | -0.39 | -1.56 | -0.95 | -0.99 |

## 4.2 Evaluation Index

In this paper, the following four evaluation functions will be used in the prediction of the chemical industry index, which are mean square error (MSE), root mean square error (RMSE), mean absolute error (MAE) and mean absolute percentage error (MAPE), these evaluation functions reflect the model fitting and prediction effects from different perspectives[40].

(1) Mean square error

The mean square error is the average of the squares of the differences between the predicted and true values. The smaller the MSE, the closer the model predicts to the true value. The formula for calculating the MSE is as follows:

$$MSE = \frac{1}{n}\sum_{i=1}^{n}(\hat{y}_i - y_i)^2 \qquad (18)$$

Where $\hat{y}_i, y_i$ denote the sample predicted value, true value, respectively, and $n$ denotes the sample size.

(2) Root mean square error

Root Mean Square Error (RMSE) is the square root of the mean square error, which converts the error into the same units as the original data. The smaller the RMSE, the closer the model predictions are to the true values. The formula for calculating RMSE is as follows:

$$RMSE = \sqrt{\frac{1}{n}\sum_{i=1}^{n}(\hat{y}_i - y_i)^2} \qquad (19)$$

Where $\hat{y}_i, y_i$ denote the sample predicted value, true value, respectively, and $n$ denotes the sample size.

(3) Average absolute error

Mean Absolute Error measures the accuracy of the prediction model by calculating the absolute error between the predicted value and the actual value and averaging it. The smaller the MAE, the closer the model predicts to the true value. The formula for calculating the MAE is as follows:

$$MAE = \frac{1}{n}\sum_{i=1}^{n}|\hat{y}_i - y_i| \qquad (20)$$

where $\hat{y}_i, y_i$ denote the sample predicted value, true value, respectively, and $n$ denotes the sample size.

(4) Mean absolute percentage error

Mean Absolute Percentage Error measures the accuracy of the prediction model by calculating the percentage error between the predicted value and the actual value and averaging it. The smaller the MAPE, the closer the model predicts to the true value. The formula for calculating MAPE is as follows:

$$MAPE = \frac{1}{n}\sum_{i=1}^{n}\left|\frac{\hat{y}_i - y_i}{y_i}\right| \times 100\% \qquad (21)$$

Where $\hat{y}_i, y_i$ denote the sample predicted value, true value, respectively, and *n* denotes the sample size.

## 4.3 Experimental Process

Due to the linear nature of the VARMA model, it is suitable for stationary or near-stationary multivariate time series. However, in non-stationary time series forecasting, the VARMA model may not effectively capture the trend of the series. In contrast, when predicting non-stationary time series, the LSTM model can handle nonlinear time series data and effectively capture long-term dependencies through its memory units and gate mechanisms. Therefore, this study divides the experimental process into stationary sequence prediction and non-stationary sequence prediction to compare the predictive performance of different models.

### 4.3.1 Stationary Series Prediction

As indicated in Table 1, the first-order differenced series are stationary. Therefore, modeling and forecasting are performed on the differenced data, followed by inverse differencing to obtain predictions for the original data and compare the predictive performance of different models.

For the VARMA and VARMAX models, the dependent variables are the indices for synthetic fibers, rubber, and plastics, and the exogenous variables are energy indices, exchange rates, and traffic indices. The Akaike Information Criterion (AIC) is utilized for model order determination. The data is split into training, validation, and test sets in a 6:2:2 ratio, and the Mean Squared Error (MSE), Root Mean Squared Error (RMSE), Mean Absolute Error (MAE), and Mean Absolute Percentage Error (MAPE) are calculated for the test set data and different forecasting periods.

For DeepVARMA-re, DeepVARMA-en, and DeepVARMA, a grid search method is employed to select appropriate model hyperparameters, with the sigmoid and ReLU functions chosen as activation functions for the gate units and memory units, respectively. Adam is selected as the optimizer, and a time step of 3 is used for predicting the chemical industry indices. Finally, the MSE, RMSE, MAE, and MAPE are calculated for the test set data and different forecasting periods.

### 4.3.2 Non-stationary Series Prediction

As indicated in Table 1, the original series are non-stationary. Based on the original data, modeling and forecasting are conducted using VARMA, VARMAX, DeepVARMA-re, DeepVARMA-en, and DeepVARMA. Additionally, LSTM, Random Forest (RF), and XGBoost are selected as comparative models from the machine learning domain. The models are used to predict the indices for synthetic fibers, rubber, and plastics, and the predictive performance of DeepVARMA-re, DeepVARMA-en, and DeepVARMA is compared with that of the statistical models VARMA and VARMAX, as well as the machine learning models LSTM, Random Forest (RF), and XGBoost.

## 4.4 Main results

This paper compares the predictive performance of seven models, namely VARMA, VARMAX, DeepVARMA-re, DeepVARMA-en, DeepVARMA, LSTM, RF, and XGBoost, for forecasting chemical industry indices. The specific results are shown in Tables 3 and 4:

Table 3 presents the predictive performance of VARMA, DeepVARMA, and other models under different stationarity conditions for the indices of synthetic fibers, plastics, and rubber. All models perform better in predicting stationary sequences compared to non-stationary sequences, and statistical models exhibit greater volatility in predicting sequences with different stationarity conditions compared to machine learning models. Table 4 displays the predictive performance of different models under various forecasting periods and time steps for non-stationary sequences. It is observed that the MSE increases with the increase in forecasting periods and time steps for all models, indicating that the predictive performance of the models is influenced by the time span and forecasting periods, with long-term forecasting performance significantly inferior to short-term forecasting performance.

**Table 3** Predictive effectiveness of the models on the test set

| experimental scenario | steady sequence | | | | non-stationary sequence | | | |
|---|---|---|---|---|---|---|---|---|
| | Chemical Fiber Index | | | | Chemical Fiber Index | | | |
| | MSE | RMSE | MAE | MAPE | MSE | RMSE | MAE | MAPE |
| VARMA | 0.064 | 0.252 | 0.171 | 0.164% | 4.977 | 2.231 | 2.941 | 2.811% |
| VARMAX | 0.064 | 0.253 | 0.165 | 0.158% | 6.340 | 2.518 | 4.758 | 4.544% |
| DeepVARMA-re | 0.022 | 0.148 | 0.097 | 0.093% | 0.061 | 0.246 | 0.171 | 0.164% |
| DeepVARMA-en | **0.001** | **0.033** | **0.042** | **0.041%** | 7.263 | 2.695 | 3.608 | 3.411% |
| DeepVARMA | 0.018 | 0.134 | 0.073 | 0.075% | **0.046** | **0.215** | **0.164** | **0.157%** |
| LSTM | - | - | - | - | 0.177 | 0.421 | 0.279 | 0.272% |
| RF | - | - | - | - | 0.493 | 0.702 | 0.475 | 0.441% |
| XGBoost | - | - | - | - | 0.149 | 0.386 | 0.253 | 0.262% |
| | Plastics Index | | | | Plastics Index | | | |
| | MSE | RMSE | MAE | MAPE | MSE | RMSE | MAE | MAPE |
| VARMA | 0.081 | 0.284 | 0.217 | 0.211% | 58.477 | 7.647 | 6.044 | 5.946% |
| VARMAX | 0.079 | 0.281 | 0.210 | 0.204% | 37.663 | 6.137 | 3.248 | 3.186% |
| DeepVARMA-re | 0.019 | 0.139 | 0.108 | 0.105% | 0.173 | 0.416 | 0.491 | 0.473% |
| DeepVARMA-en | **0.003** | **0.057** | **0.084** | **0.082%** | 19.044 | 4.364 | 4.466 | 4.402% |
| DeepVARMA | 0.016 | 0.128 | 0.094 | 0.096% | **0.112** | **0.335** | **0.309** | **0.302%** |
| LSTM | - | - | - | - | 0.766 | 0.875 | 0.538 | 0.503% |
| RF | - | - | - | - | 3.244 | 1.801 | 0.938 | 0.762% |
| XGBoost | - | - | - | - | 0.347 | 0.589 | 0.505 | 0.487% |

|  | Rubber Index | | | | Rubber Index | | | |
| --- | --- | --- | --- | --- | --- | --- | --- | --- |
|  | MSE | RMSE | MAE | MAPE | MSE | RMSE | MAE | MAPE |
| VARMA | 0.308 | 0.555 | 0.379 | 0.319% | 41.835 | 6.468 | 7.376 | 6.084% |
| VARMAX | 0.285 | 0.534 | 0.361 | 0.307% | 35.058 | 5.921 | 6.629 | 5.813% |
| DeepVARMA-re | 0.100 | 0.317 | 0.219 | 0.185% | 0.570 | 0.755 | 0.473 | 0.382% |
| DeepVARMA-en | **0.022** | **0.149** | **0.066** | **0.056%** | 32.753 | 5.723 | 5.986 | 5.273% |
| DeepVARMA | 0.106 | 0.325 | 0.185 | 0.118% | 0.388 | 0.623 | **0.424** | **0.356%** |
| LSTM | - | - | - | - | 0.440 | 0.663 | 0.569 | 0.486% |
| RF | - | - | - | - | 2.599 | 1.612 | 1.038 | 0.821% |
| XGBoost | - | - | - | - | **0.379** | **0.616** | 0.634 | 0.548% |

In stationary sequence prediction, DeepVARMA-en performs the best, followed by DeepVARMA. This might be attributed to the fact that DeepVARMA-en directly encodes exogenous variables using LSTM and integrates the encoded features into the VARMAX model for predicting the original endogenous variables. Throughout the process, the information of exogenous variables is more directly integrated into the prediction model, possibly enabling a more direct capture of the influence of exogenous variables on endogenous variables. On the other hand, DeepVARMA first predicts the original sequence using LSTM, then encodes the residual with exogenous variables, and finally predicts the residual using VARMAX. As the original sequence is stationary, the information contained in the residuals may not accurately reflect the part of the original sequence that was not captured by LSTM, introducing some additional complexity.

In non-stationary sequence prediction, DeepVARMA performs the best, followed by DeepVARMA-re. The possible reason lies in the capability of the LSTM prediction layer in DeepVARMA to effectively capture the long-term dependencies and potential non-linear features present in the original sequence. Additionally, by encoding exogenous variables with LSTM and extracting useful features for residual prediction, DeepVARMA can utilize the information contained in exogenous variables to enhance prediction accuracy further. Compared to LSTM, RF, and XGBoost, DeepVARMA not only utilizes the predictive ability of LSTM on the original sequence but also leverages the VARMAX prediction layer to capture the correlations and complex dynamic relationships among multiple variables in multivariate time series data. This dual prediction mechanism enhances the robustness of the prediction results. Figure 8 illustrates the fitting effect of DeepVARMA on the testing dataset for the indices of synthetic fibers, plastics, and rubber. It clearly shows a high degree of consistency between the true and predicted sequences, indicating DeepVARMA's effective ability to capture and handle non-stationary sequences.

**Table 4** Mean square error of the models in predicting the chemical industry index under non-stationary series with different number of forecasting periods

| | Chemical Fiber Index | | | | | | | |
|---|---|---|---|---|---|---|---|---|
| | 1 | 5 | 10 | 15 | 1:5 | 1:10 | 1:15 | 1:20 |
| VARMA | 11.903 | 11.381 | 11.493 | 12.748 | 12.615 | 13.613 | 16.873 | 15.131 |
| VARMAX | 10.394 | 11.473 | 11.874 | 11.579 | 11.235 | 12.513 | 11.983 | 13.631 |
| DeepVARMA-re | 0.395 | 0.546 | 0.791 | 0.984 | **0.328** | 0.565 | 0.553 | 0.672 |
| DeepVARMA-en | 11.795 | 11.483 | 12.188 | 11.493 | 10.492 | 9.284 | 9.183 | 10.462 |
| DeepVARMA | **0.134** | **0.475** | **0.626** | **0.931** | 0.331 | **0.433** | **0.530** | **0.641** |
| LSTM | 1.594 | 1.289 | 1.376 | 2.472 | 2.483 | 2.857 | 3.183 | 4.879 |
| RF | 2.397 | 2.164 | 2.873 | 3.283 | 3.834 | 3.982 | 4.781 | 5.736 |
| XGBoost | 1.473 | 3.293 | 2.573 | 2.789 | 3.102 | 2.395 | 1.583 | 3.284 |
| | Plastics Index | | | | | | | |
| | 1 | 5 | 10 | 15 | 1:5 | 1:10 | 1:15 | 1:20 |
| VARMA | 16.685 | 17.483 | 15.284 | 17.583 | 112.301 | 100.395 | 104.212 | 120.265 |
| VARMAX | 15.573 | 16.385 | 14.291 | 15.372 | 92.578 | 84.832 | 72.583 | 68.329 |
| DeepVARMA-re | 1.382 | 1.284 | 0.974 | 1.682 | 2.454 | 1.677 | **1.357** | 2.940 |
| DeepVARMA-en | 12.684 | 11.283 | 12.785 | 13.395 | 38.828 | 40.260 | 41.970 | 43.269 |
| DeepVARMA | **0.982** | **1.548** | **0.832** | **1.287** | **1.244** | **1.437** | 1.557 | **2.580** |
| LSTM | 2.439 | 2.194 | 2.785 | 3.248 | 6.843 | 7.593 | 10.394 | 11.492 |
| RF | 4.695 | 3.685 | 5.892 | 5.381 | 11.223 | 18.444 | 19.239 | 24.086 |
| XGBoost | 1.493 | 2.694 | 3.854 | 3.695 | 5.385 | 6.732 | 9.578 | 10.485 |
| | Rubber Index | | | | | | | |
| | 1 | 5 | 10 | 15 | 1:5 | 1:10 | 1:15 | 1:20 |
| VARMA | 15.694 | 16.293 | 16.876 | 17.484 | 121.338 | 133.608 | 135.657 | 132.205 |
| VARMAX | 15.548 | 13.462 | 14.348 | 15.285 | 134.725 | 123.524 | 115.819 | 127.474 |
| DeepVARMA-re | 1.164 | 1.238 | 1.834 | 2.472 | 1.689 | 2.545 | 1.729 | 2.640 |
| DeepVARMA-en | 13.573 | 13.896 | 14.385 | 16.483 | 64.925 | 78.377 | 85.282 | 90.538 |
| DeepVARMA | **1.042** | **0.843** | **1.382** | **1.573** | **1.499** | **1.315** | **1.469** | **2.430** |
| LSTM | 2.943 | 3.458 | 5.283 | 5.893 | 7.493 | 7.392 | 10.312 | 12.481 |
| RF | 4.343 | 5.392 | 5.983 | 6.493 | 13.684 | 19.483 | 22.584 | 25.495 |
| XGBoost | 1.495 | 3.684 | 2.329 | 3.183 | 1.723 | 2.274 | 3.483 | 3.492 |

In the prediction process of non-stationary sequences under different forecast horizons and time steps, DeepVARMA demonstrates the best performance, followed by DeepVARMA-re. Compared to DeepVARMA-en as well as the statistical models VARMA and VARMAX, machine learning models such as LSTM, RF, and XGBoost perform better in non-stationary sequences.

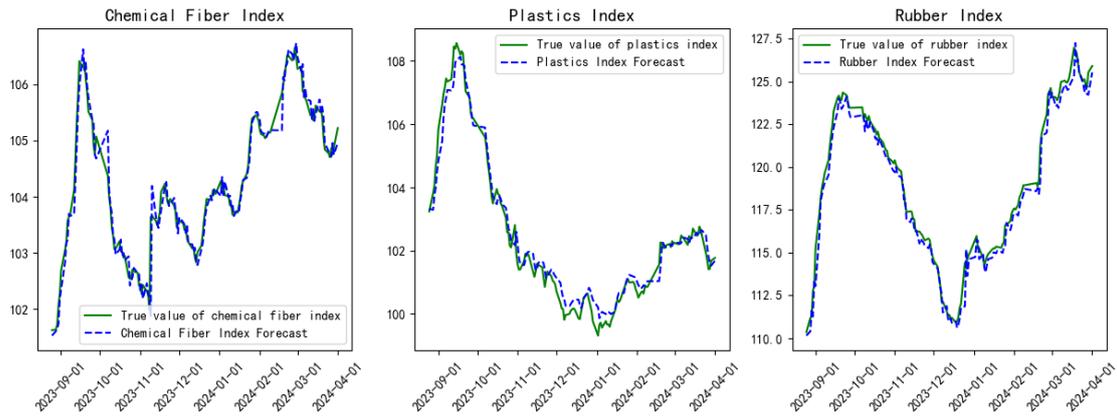

**Fig. 8** Effectiveness of fitting the test set of DeepVARMA models

## 5 Conclusion

In this study, we addressed the prediction problem of composite material indices in the chemical industry by designing forecasting algorithms for DeepVARMA, DeepVARMA-re, and DeepVARMA-en. These algorithms were compared with traditional VARMA and VARMAX models, as well as machine learning models including LSTM, RF, and XGBoost. Through a series of experiments and analyses, the following conclusions were drawn:

Firstly, in the prediction of stationary sequences, DeepVARMA-en achieved optimal prediction accuracy by directly combining the encoding of exogenous variables using LSTM and the prediction of original endogenous variables using the VARMAX model. In the prediction of non-stationary sequences, DeepVARMA adopted a staged approach. It initially utilized LSTM for preliminary prediction and residual calculation on the original sequence. Subsequently, it encoded exogenous variables using LSTM and employed the VARMAX model to predict the residuals. Finally, the results from both stages were combined to obtain the final prediction. This strategy demonstrated higher adaptability and accuracy in handling non-stationary sequences. Furthermore, compared to predictions of stationary sequences, traditional VARMA and VARMAX models exhibited greater fluctuations in predicting non-stationary sequences. In contrast, DeepVARMA demonstrated greater flexibility and robustness in predicting chemical industry indices. This can be attributed to the limitations of VARMA models in capturing nonlinear and non-stationary characteristics, whereas LSTM-based algorithms excel in handling complex sequences. The introduction of the VARMAX model further enhanced prediction stability. Additionally, compared to machine learning models such as LSTM, RF, and XGBoost, DeepVARMA exhibited unique advantages in predicting chemical industry indices. It not only leveraged LSTM's predictive ability on original sequences but also enabled the VARMAX prediction layer to capture correlations and complex dynamic relationships among multiple variables in multivariate time series data, which machine learning models cannot capture.

The proposed DeepVARMA prediction algorithm achieved significant results in predicting chemical industry indices. By comparing with traditional and machine learning models, this study further validated the effectiveness and superiority of the proposed algorithm, providing a powerful tool for market analysis and investment decision-making in the chemical industry. However, while the model performs well in forecasting the chemical industry index, its applicability may be constrained by industry-specific characteristics. Enhancing the model's generalization capability, robustness, and interpretability constitutes our future research direction.

## Acknowledgments

This work is supported by the Late Subsidized Project of National Social Science Foundation of China (22FGLB056), the National Natural Science Foundation of China Youth Science Foundation Project (71701223), the National Statistical Science Research Project (2023LY078), the Seventh Batch of Young Scientific Research and Innovation Team of Central University of Finance and Economics.